\documentclass[runningheads,a4paper]{llncs}

\usepackage{times}

\usepackage{amssymb}
\usepackage{url}
\usepackage{graphicx}
\usepackage{booktabs}
\usepackage{pslatex}

\urldef{\mailsa}\path|{philipp.bucher, bernard.moret}@epfl.ch|
\newcommand{\keywords}[1]{\par\addvspace\baselineskip
\noindent\keywordname\enspace\ignorespaces#1}
\begin{document}

\mainmatter  % start of an individual contribution

\title{Phylogenetic Analysis of Cell Types using Histone Modifications}

\titlerunning{Phylogenetic Analysis of Cell Types using Histone Modifications}

\author{Nishanth Ulhas Nair$^{1,\#}$ \and Yu Lin$^{1}$\and Philipp Bucher$^{2,*}$\and Bernard M.E. Moret$^{1,*}$}
\authorrunning{N.U. Nair, Y. Lin, P. Bucher, and B.M.E. Moret}

\institute{${}^1$ School of Computer and Communication Sciences.\\
${}^2$ School of Life Sciences. \\
\'{E}cole Polytechnique F\'{e}d\'{e}rale de Lausanne (EPFL), Lausanne, Switzerland. \\
$\#$ NUN's project was funded by Swiss National Science Foundation.\\
$*$ corresponding authors \\
\mailsa\\
}
\maketitle

\begin{abstract}
  In cell differentiation, a cell of a less specialized type becomes one of a
  more specialized type, even though all cells have the same genome.
  Transcription factors and epigenetic marks like histone modifications can
  play a significant role in the differentiation process.  In this paper, we
  present a simple analysis of cell types and differentiation paths using
  phylogenetic inference based on ChIP-Seq histone modification data.  We
  propose new data representation techniques and new distance measures for
  ChIP-Seq data and use these together with standard phylogenetic inference
  methods to build biologically meaningful trees that indicate how diverse
  types of cells are related.  We demonstrate our approach on H3K4me3 and
  H3K27me3 data for 37 and 13 types of cells respectively, using the dataset to explore various
  issues surrounding replicate data, variability between cells of the same
  type, and robustness.  The promising results we obtain point the way to a new
  approach to the study of cell differentiation. 

\end{abstract}

\section{Introduction and Background}
In developmental biology, the process by which a less specialized cell becomes
a more specialized cell type is called cell differentiation.  Since all cells
in one individual organism have the same genome, epigenetic factors and
transcriptional factors play an important role in cell differentiation
\cite{lee2004histone,lister2011hotspots,lobe1992transcription}.
Thus a study of epigenetic changes among different cell types
is necessary to understand cell development.

Histone modifications form one important class of epigenetic marks;
such modifications
have been found to vary across various cell types and to play a role in gene
regulation~\cite{berger2002histone}.  Histones are proteins that package DNA
into structural units called nucleosomes \cite{nelson2010lehninger}.  These
histones are subject to various types of modifications (methylation,
acetylation, phosphorylation, and ubiquitination), modifications that alter
their interaction with DNA and nuclear proteins. In turn, changes in these
interactions influence gene transcription and genomic function.
In the last several years a high-throughput, low-cost, sequencing technology
called ChIP-Seq has been used in capturing these histone marks
on a genome-wide scale \cite{barski2007high,mardis2007chip}.
A study of how histone marks change across various cell types could
play an important role in our understanding of developmental biology
and how cell differentiation occurs, particularly as the epigenetic
state of chromatin is inheritable across cell generations~\cite{inheritance}.

Since cell differentiation transforms less specialized cell types into more
specialized ones and since most specialized cells of one organ cannot be
converted into specialized cells of some other organ, the paths of
differentiation together form a tree, in many ways similar to the phylogenetic
trees used to represent evolutionary histories.  In evolution, present-day
species have evolved from some ancestral species, while in cell development the
more specialized cells have evolved from less specialized cells.  Moreover,
observed changes in the epigenetic state are inheritable, again much as
mutations in the genome are (although, of course, through very different
mechanisms and at very different scales); and in further similarity, epigenetic
traits are subject to stochastic changes, much as in genetic mutations.
(It should be noted that we are interested here in populations of cells of
a certain type, not all coming from the same individual, rather than in
developmental lineages of cells within one individual.)\ \ 
Finally, one may object that derived and more basic cell types coexist
within the body, while phylogenetic analysis places all modern data at
the leaves of the tree and typically qualifies internal nodes as
``ancestral".\ \ However, species in a phylogenetic tree correspond to
paths, not to nodes.  In particular, a species that has survived millions
of years until today and yet has given rise to daughter species,
much like a basic cell type that is observed within the organism, but from
which derived cell types have also been produced and observed,
is simply a path to a leaf in the tree, a path along which
changes are slight enough not to cause a change in identification.
(The time scale makes such occurrences unlikely in the case of
species phylogenies, but the framework is general enough to include them.)

Therefore it may be
possible to use or adapt some of the techniques used in building phylogenetic
trees for building \emph{cell-type trees}.
There are of course significant differences between a phylogenetic tree
and a cell-type tree. Two major differences stand out.
The more significant difference is the lack of well
established models for changes to histone marks during cell differentiation,
as compared to the DNA and amino-acid mutation models in common usage
in research in molecular evolution.   The other difference is that functional
changes in cell differentiation are primarily driven by programmed mutational
events rather than by selection---and this of course makes it all the harder
to design a good model.  In spite of these differences, we felt that
phylogenetic approaches could be adapted to the analysis of cell
differentiation.

In this paper, we provide evidence that such a scenario is possible.  We do
this by proposing new data representation techniques and distance measures,
then by applying standard phylogenetic methods to produce biologically
meaningful results.  We used data on two histone
modifications (but mostly on H3K4me3) for 37 cell types, including replicate
data, to construct cell-type trees---to our knowledge, these are the
first such trees produced by computational methods.
We show that preprocessing the data is very important: not only are ChIP-Seq
data fairly noisy, but the ENCODE data are based on several individuals
and thus adds an independent source of noise.  We also outline some of the
computational challenges in the analysis of cell differentiation, opening new
perspectives that may prove of interest to computer scientists, biologists, and
bioinformaticians. 

\section{Methods}

\subsection{Model of differentiation for histone marks}
We assume that histone marks can be independently gained or lost in regions of
the genome as cells differentiate from a less specialized type to a more
specialized one.  Histones marks are known to disappear from less specialized
cell types or to appear in more specialized ones and are often correlated with
gene expression, so our assumption is reasonable.
The independence assumption simply reflects our lack of knowledge, but
it also enormously simplifies computations.

\subsection{Data representation techniques}
The analysis of ChIP-Seq data typically starts with a peak-finding step
that defines a set of chromosomal regions enriched in the target molecule.
We therefore use peak lists as the raw data for our study.  We can decide
on the presence or absence of peaks at any given position and treat this
as a binary character, matching our model of gain or loss of histone marks.
Since all of the cell types have the same genome (subject only to individual
SNPs or varying copy numbers), we can compare specific regions across cell
types.   Therefore we code the data into a matrix in which each row is
associated with a different ChIP-Seq library (a different cell type or
replicate), while each column is associated with a specific genomic region.

We use two different data representations for the peak data for each cell type.
Our first method is a simple windowing (or binning) method.  We divide
the genome into bins of certain sizes; if the bin contains at least one
peak, we code it 1, otherwise we code it~0.  The coding of each library
is thus independent of that of any other library.

Our second method uses overlap and takes into account all libraries at
once.  We first find interesting regions in the genome, based on peaks.
Denote the $i$th peak in library $n$ as $P_i^n = [P_{iL}^n,P_{iR}^n]$, where
$P_{iL}^n$ and $P_{iR}^n$ are the left and right endpoints (as basepair
indices).  Consider each peak as an interval on the genome (or on the
real line) and build the \emph{interval graph} defined by all peaks in all
libraries.  An interval graph has one vertex for each interval and
an edge between two vertices whenever the two corresponding intervals
overlap~\cite{fishburn1985interval}.  We simply want the connected
components of the interval graph.
\begin{definition}
  An interval in the genome is an \emph{interesting region} iff it
  corresponds to a connected component of the interval graph.
%  (see Fig.~\ref{IR-fig}).
\end{definition}
%\begin{figure}[t]
%  \centerline{\includegraphics[height=1.5in]{IR_region.eps}}%
%  \caption{The interval graph coding of peaks.  Black horizontal bars denote
%  peak regions.}%
%  \label{IR-fig}%
%\end{figure}%
Finding these interesting regions is straightforward.
Choose a chromosome, let $PS$ be its set of peaks, set $AS=\{\varnothing\}$
and $z = 0$, and enter the following loop:\\
\begin{enumerate}
  \itemsep 1pt
  \item
    $P_{i^*}^{*} = \arg\min_{P_i^n \in PS} P_{iL}^n$.
    Set $a=P_{i^{*} L}^{*}$ and $AS = AS \cup \{P_{i^*}^{*}\}$
  \item
    Set
    $S = \{P\mid P\cap P_{i^*}^{*}\neq\varnothing\textnormal{ and }P\in PS\}$
    and $AS = AS\cup S$.
  \item
    If $S$ is not empty, then find
    $P_{i^*}^{*} = \arg\max_{P_i^n\in PS} P_{iR}^{n}$ and go to step 2.
  \item
    Let $b = P_{i^{*}R}^{*}$ and set $PS = PS - AS$.
  \item
    The interesting region lies between $a$ and $b$, $IR[a,b]$.
    Let $D_{IR}^{n}[z]$ be the data representation for $IR[a,b]$
    in library $n$. Set $z = z + 1$.
    Set $D_{IR}^{n}[z] = 1$ if there is a peak in library $n$ that lies
    in $IR[a,b]$; otherwise set $D_{IR}^{n}[z] = 0$ ($1 \leq n \leq N$).
\end{enumerate}
Repeat this procedure for all chromosomes in the genome.  The algorithm
takes time linear in the size of the genome to identify the interesting regions. 

For a given collection of libraries, these interesting regions have a unique
representation.  
We assume that it is in these interesting regions that histone marks are lost
or gained and we consider that the size of the histone mark (which depends at
least in part on the experimental procedures and is typically noisy) does
not matter.  Our major reason for this choice of representation is noise
elimination: since the positioning of peaks and the signal strength both vary
from cell to cell as well as from test to test, we gain significant robustness
(at the expense of detail) by merging all overlapping peaks into one signal,
which we use to decide on the value of a single bit. The loss of information
may be illusory (because of the noise), but in any case we do not need a lot
of information to build a phylogeny on a few dozen cell types.

\subsection{Phylogenetic analysis}
Phylogenetic analysis attempts to infer the evolutionary relationships
of modern species or \emph{taxa}---they could also be proteins, binding sites,
regulatory networks, etc.  The best tools for phylogenetic inference,
based on maximum parsimony (MP) or maximum likelihood (ML), use established
models of sequence evolution, something for which we have no equivalent in
the context of cell differentiation.  However, one class of phylogenetic
inference methods uses variations on clustering, by computing measures of
distance (or similarity) to construct a hierarchical clustering
that is assimilated to a phylogenetic tree.  This type of method is
applicable to our problem, provided we can define a reasonable measure
of distance, or similarity, between cell types in terms of our data
representations.  (We are not implying that models of differentiation
do not exist nor that they could not be derived, but simply stating that
none exist at present that could plausibly be used for maximum-likelihood
phylogenetic inference.)\ \ Finally more that, with 0/1 data, we can also
use an MP method, in spite of the absence of a valid model
of character evolution.

In a cell-type tree, most cell types coexist in the present; thus at least some
of them can be found both at leaves and at internal nodes.  (We are unlikely to
have data for all internal nodes, as we cannot claim to have observed all cell
types.)\ \ Fortunately, phylogenetic inference still works in such cases: as
mentioned earlier, when the same taxon should be associated with both a leaf
and an internal node, we should simply observe that each edge on the path from
that internal node to that leaf is extremely short, since that distance between
the two nodes should be zero (within noise limits).  The tree inferred will
have the correct shape; however, should we desire to reconstruct the basic
cell types, then we would have to \emph{lift} some of the leaf data by copying
them to some internal nodes.

From among the distance-based methods, we chose to use the most commonly
used one, Neighbor-Joining (NJ)~\cite{saitou1987neighbor}.
While faster and possibly better distance-based methods exist, such
as FastME~\cite{fastme}, it was not clear that their advantages
would still obtain in this new domain; and, while very simple, the NJ
method has the advantage of not assuming a constant rate of evolution
across lineages. 
In each of the two data representation approaches, we compute pairwise distance
between two libraries as the Hamming distance of their representations.
(The Hamming distance between two strings of equal length is the number of
positions at which corresponding symbols differ.)\ \ 
We thus obtain a distance matrix between all pairs of histone modification
libraries; running NJ on this matrix yields an unrooted tree. 
For MP, we used the TNT software~\cite{TNT}.

\subsection{On the inference of ancestral nodes}
We mentioned that lifting some of the leaf data into internal nodes is
the natural next step after tree inference.  However, in general,
not all internal nodes can be labelled in this way, due mostly
to sampling issues: we may be missing the type that should be associated
with a particular internal node, or we may be missing enough fully
differentiated types that some internal tree nodes do not correspond to
any real cell type.  Thus we are faced with a problem of ancestral
reconstruction and, more specifically, with three distinct questions:
\begin{itemize}
  \itemsep 1pt
  \item
    For a given internal node, is there a natural lifting from a leaf?
  \item
    If there is no suitable lifting, is the node nevertheless a natural
    ancestor---i.e., does it correspond to a valid cell type?
  \item
    If the node has no suitable lifting and does correspond to a valid
    cell type, can we infer its data representation?
\end{itemize}
These are hard questions, in terms of both modelling and computational
complexity; they are further complicated by the noisy nature of the data.
Such questions remain poorly solved in standard phylogenetic analysis;
in the case of cell-type trees, we judged it best not to
address these problems until the tree inference part is better understood and
more data are analyzed.

\section{Experimental Design}
The histone modification ChIP-Seq data were taken from the ENCODE project
database (UW ENCODE group) for human (hg19) data \cite{encode2011user}.  We
carried out experiments on both H3K4me3 and H3K27me3 histone mark data.
H3K4me3 is a well studied histone mark usually associated with gene activation,
while the less well studied H3K27me3 is usually associated with gene
repression~\cite{mikkelsen2007genome}.  We used data for cell
types classified as ``normal'' and for embryonic stem cells---we did not
retain cancerous or EBV cells as their differentiation processes might
be completely distinct from those of normal cells.
The ENCODE project provides peaks of ChIP-Seq data for each replicate of each
cell type.  We therefore used their peaks as the raw input data for our
work.  For the windowing representation, we used bins of 200 bp: this is a good
size for histone marks, because 147 bp of DNA wrap around the histone and
linker DNA of about 80 bp connect two histones, so that each bin
represents approximately the absence or presence of just one histone
modification.  We programmed our procedures in $R$ and used the NJ
implementation from the {\it ape} library in $R$.

Table \ref{tab:celltypes}
\begin{table}[t]% 
  \caption{\footnotesize\it
   Cell names,  short description, and general group for H3K4me3 data.
   For details see the ENCODE website \cite{encode_website}.  
   }%
  \footnotesize
  \centerline{\begin{tabular}{lll}%
     \toprule
     Cell Name & Short Description & Group \\
     \midrule
     AG04449  & fetal buttock/thigh fibroblast & Fibroblast  \\ 
     AG04450 & fetal lung fibroblast & Fibroblast \\
     AG09319  & gum tissue fibroblasts & Fibroblast  \\
     AoAF   & aortic adventitial fibroblast cells & Fibroblast  \\
     BJ   & skin fibroblast & Fibroblast   \\
     CD14  & Monocytes-CD14+ from human leukapheresis production & Blood \\
     CD20(1)   &  B cells replicate, African American & Blood \\
     CD20(2) and CD20(3) & B cells replicates, Caucasian & Blood \\
     hESC & undifferentiated embryonic stem cells & hESC  \\
     HAc   & astrocytes-cerebellar  & Astrocytes \\
     HAsp   & astrocytes spinal cord  & Astrocytes \\
     HBMEC   &  brain microvascular endothelial cells  & Endothelial \\
     HCFaa   & cardiac fibroblasts- adult atrial & Fibroblast  \\
     HCF   &  cardiac fibroblasts & Fibroblast  \\
     HCM   & cardiac myocytes  & Myocytes \\
     HCPEpiC   &  choroid plexus epithelial cells & Epithelial  \\
     HEEpiC   & esophageal epithelial cells & Epithelial \\
     HFF   & foreskin fibroblast & Fibroblast  \\
     HFF MyC   & foreskin fibroblast cells expressing canine cMyc & Fibroblast\\
     HMEC   & mammary epithelial cells & Epithelial \\
     HPAF   &  pulmonary artery fibroblasts & Fibroblast \\
     HPF   & pulmonary fibroblasts isolated from lung tissue & Fibroblast \\
     HRE   &  renal epithelial cells & Epithelial \\
     HRPEpiC   &  retinal pigment epithelial cells & Epithelial \\
     HUVEC   & umbilical vein endothelial cells  & Endothelial  \\
     HVMF   &  villous mesenchymal fibroblast cells & Fibroblast \\
     NHDF Neo   &  neonatal dermal fibroblasts & Fibroblast  \\
     NHEK   &  epidermal keratinocytes & Epithelial \\
     NHLF   & lung fibroblasts & Fibroblast  \\
     RPTEC   &  renal proximal tubule epithelial cells & Epithelial \\
     SAEC   & small airway epithelial cells  & Epithelial \\
     SKMC   & skeletal muscle cells  & Skeletal Muscle \\
     WI 38   & embryonic lung fibroblast cells & Fibroblast\\
     \bottomrule
    \end{tabular}}%
  \label{tab:celltypes}%
\end{table}%
show the list of the 37 cell types used for H3K4me3
data, giving for each an abbreviation and a short description.  In addition, 
the cells are classified into various groups whose names are based on their
cell type.  Keratinocytes (NHEK) is included in the Epithelial group.
We have two replicates for most cell types, but only one replicate for
types HCFaa, HFF, and CD14, and three replicates for CD20.
(CD20(1) is a B-cell from an  African-American individual
while CD20(2) and CD20(3) are from a Caucasian individual).
The replicates are biological replicates, i.e., the data come from
two independent samples.  For human Embryonic Stem Cells (hESC) we have
data for different days of the cell culture, so we shall use hESC D2
to mean data for hESC cells on day 2.  For each cell type, we shall mention
the replicate number in brackets, unless the cell type has only one replicate.

\section{Results/Discussion}

\subsection{H3K4me3 data on individual replicates}
We report on our analyses using peak data from the ENCODE database for
H3K4me3 histone modifications.  We carried out the same analyses using
H3K27me3 data, but results were very similar and so are not detailed here---we
simply give one tree for comparison purposes.
The similarity of results between the two datasets reinforces our contention
that phylogenetic analyses yield biologically meaningful results on such data.
We color-code trees to reflect the major groupings
listed in Table~\ref{tab:celltypes}.

\begin{figure*}[b]%
  \hbox to\columnwidth{%
   \includegraphics[width = 6cm]{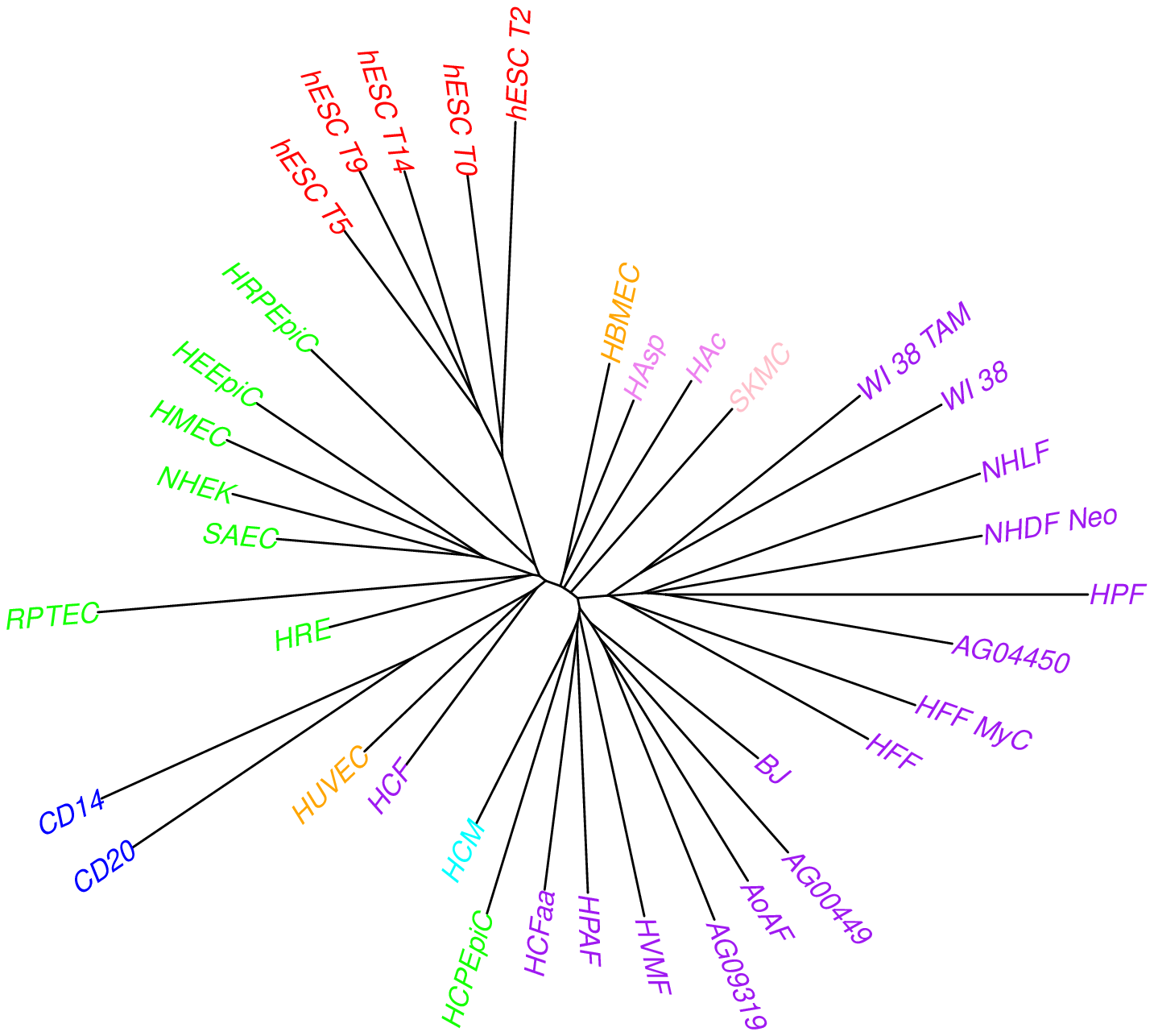}\hfil
   \includegraphics[width = 6.5cm]{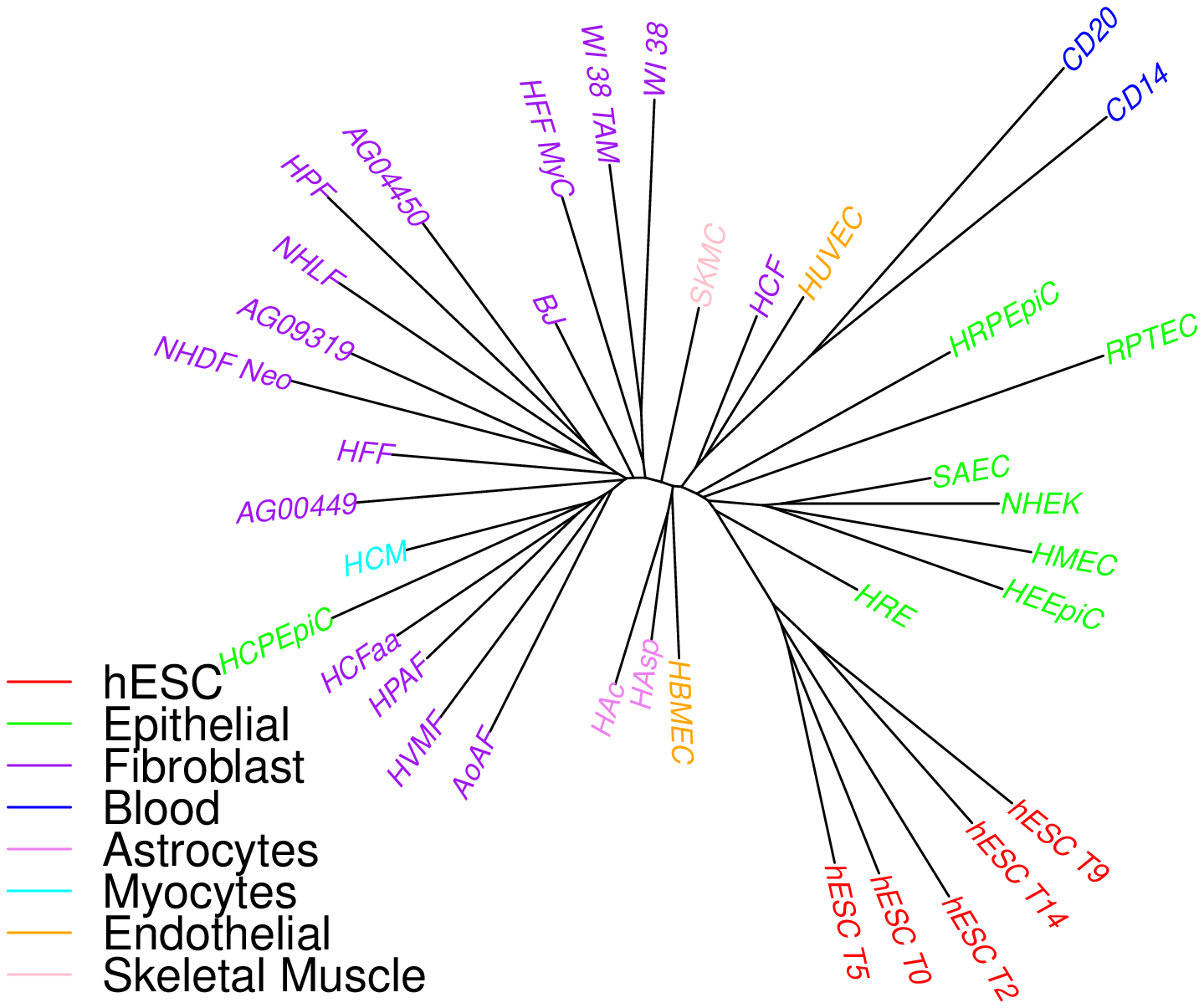}}%
  \hbox to\columnwidth{%
	  \hbox to7cm{\hfil\scriptsize\it (a) \hfil}\hfil
  	  \hbox to7cm{\hfil\scriptsize\it (b) \hfil}}%
    \caption{\footnotesize\it Cell-type tree on H3K4me3 data using only one replicate: (a) windowing representation, (b) overlap representation.}%
    \label{fig:H3K4me3_rep1}%
\end{figure*}

Fig.~\ref{fig:H3K4me3_rep1}
shows the trees constructed using only one replicate for each cell type
using both windowing and overlap representations.
The color-coding shows that embryonic stem cells and blood cells are in well
separated clades of their own, while fibroblasts and epithelial cells fall
in just two clades each.  Even within the hESC group we see that day 0
is far off from day 14 compared to its distance from day 2.
Thus epigenetic data such as histone marks do contain a lot of information
about cell differentiation history.

In order to quantify the quality of the groupings, we compute the total
number of cells in a subtree that belong to one group.  Since our groups are
based on cell type only, there could be many subdivisions possible within
each group.  Therefore we choose the two largest
such subtrees available for each group such that each subtree contains only
the leaf nodes of that group. The results are shown in Table \ref{tab:group}:
most of the cell types in each group do cluster together in the tree.  
\begin{table}[t]% 
  \caption{\footnotesize\it
  {\bf Statistics for cell-type trees on H3K4me3 data.} 2nd to 9th columns show the number of cells (of the same type) belonging to
  the largest and second-largest clades; the total number of cells of that
  type is in the top row.  Rows correspond to various methods
  (WM: windowing; OP: overlap; TP: top peaks).
  The last column contains the percent deviation ($PD$) of the
  distances between the leaves found using the NJ tree from the
  Hamming distance between the leaves.}%
\scriptsize
  \centerline{%
    \begin{tabular}{rcccccccccc}%
      \toprule
      & hESC & Epithelial & Fibroblast & Blood & Astrocytes & Myocytes & Endothelial & Skeletal Muscle & $SR$ & $PD$\\  
                             &(5)&(8)&(16)&(2)&(2)&(1)&(2)&(1)&    &(\%)\\
      \midrule
      WM (one replicate)     &5,0&6,1&8,4 &2,0&1,1&1,0&1,1&1,0&0.93&3.20\\
      OM (one replicate)     &5,0&4,1&6,3 &2,0&2,0&1,0&1,1&1,0&0.92&3.94\\
      WM (all replicates)    &5,0&6,1&11,2&2,0&1,1&1,0&1,1&1,0&0.84&3.30\\
      OM (all replicates)    &5,0&4,2&9,4 &2,0&2,0&1,0&1,1&1,0&0.78&3.88\\
      WM (all replicates)-TP &5,0&6,1&7,4 &2,0&1,1&1,0&1,1&1,0&0.81&3.73\\
      OM (all replicates)-TP &5,0&4,3&8,5 &2,0&2,0&1,0&1,1&1,0&0.74&3.98\\
      \bottomrule
    \end{tabular}}%
  \label{tab:group}%
\end{table}
Fig.~\ref{fig:H3K4me3_rep1} shows long edges
between (most) leaf nodes and their parents---a disquieting feature, as
it casts doubt as to the robustness of the tree, parts of which
could be assimilated to stars.
To quantify this observation, we measured
the $SR$ ratio, defined as $SR = \frac{\sum_{e\in I} l(e)}{\sum_{e\in E} l(e)}$,
where $I$ is the set of all edges connecting leaf nodes to their
parents, $E$ is the set of all edges in the tree, and $l(e)$ is the length of
edge $e$. If this ratio $SR$ is close to 1, then the tree looks star-shaped
with long branches to the leaves.  This ratio was 0.93 using the
windowing representation; using the overlap representation reduced it
very slightly to 0.92.  These long branches are due in part to the very
high level of noise in the data, explaining why the overlap representation
provided a slight improvement.

As a final entry in the table, we added added another measure on the tree
and the data.  The NJ algorithm is known to return the ``correct" tree when
the distance matrix is ultrametric; the technical definition does not matter
so much here as the consequence: if the matrix is ultrametric, then the
sum of the length of the edges on the path between two leaves always equals
the pairwise distance between those two leaves in the matrix.
Thus one way to estimate how far the distance matrix deviates from this
ideal is to compare its distances to the length of the leaf-to-leaf
paths in the tree:
  $$PD = \frac{\sum_{i,j} |NJ(i,j) -  M(i,j)|}{\sum_{i,j}NJ(i,j)}$$
where $i$ and $j$ are leaf nodes, $NJ(i,j)$ is the tree distance
between $i$ and $j$, and $M(i,j)$ is the matrix distance between $i$ and $j$.
A high value of $PD$ indicates that the data representations and
measures do not fit well to any tree.  We get very low values (of less than
4\% for both windowing and overlap representations), suggesting that the
distances we compute are in fact representative of a tree and thus offering
confirmation of the validity of the inference.

\subsection{H3K4me3 data with all replicates}
\begin{figure*}[t]%
  \hbox to\textwidth{%
   \includegraphics[width = 6.75cm]{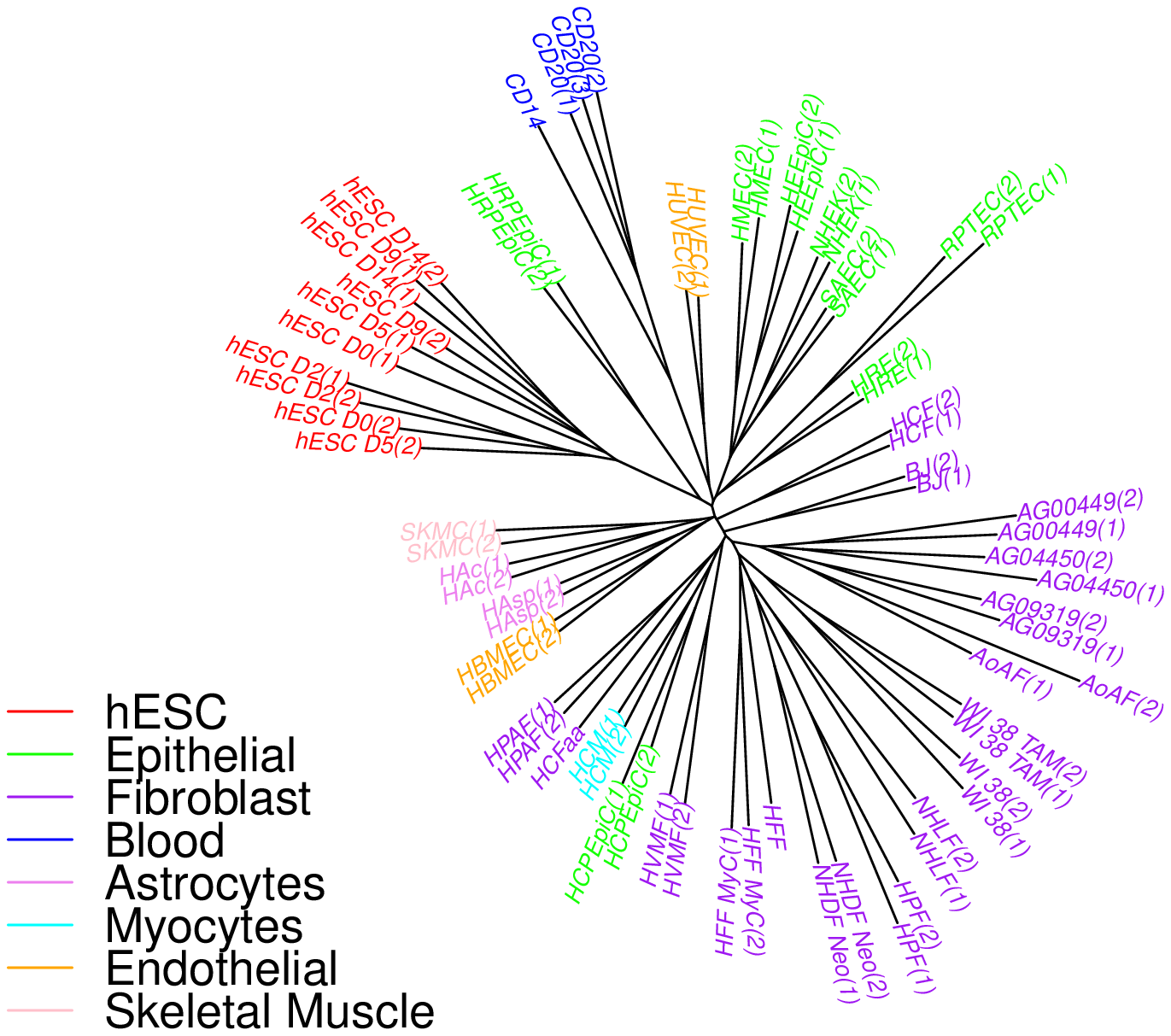}\hfil%
   \includegraphics[width = 6.2cm]{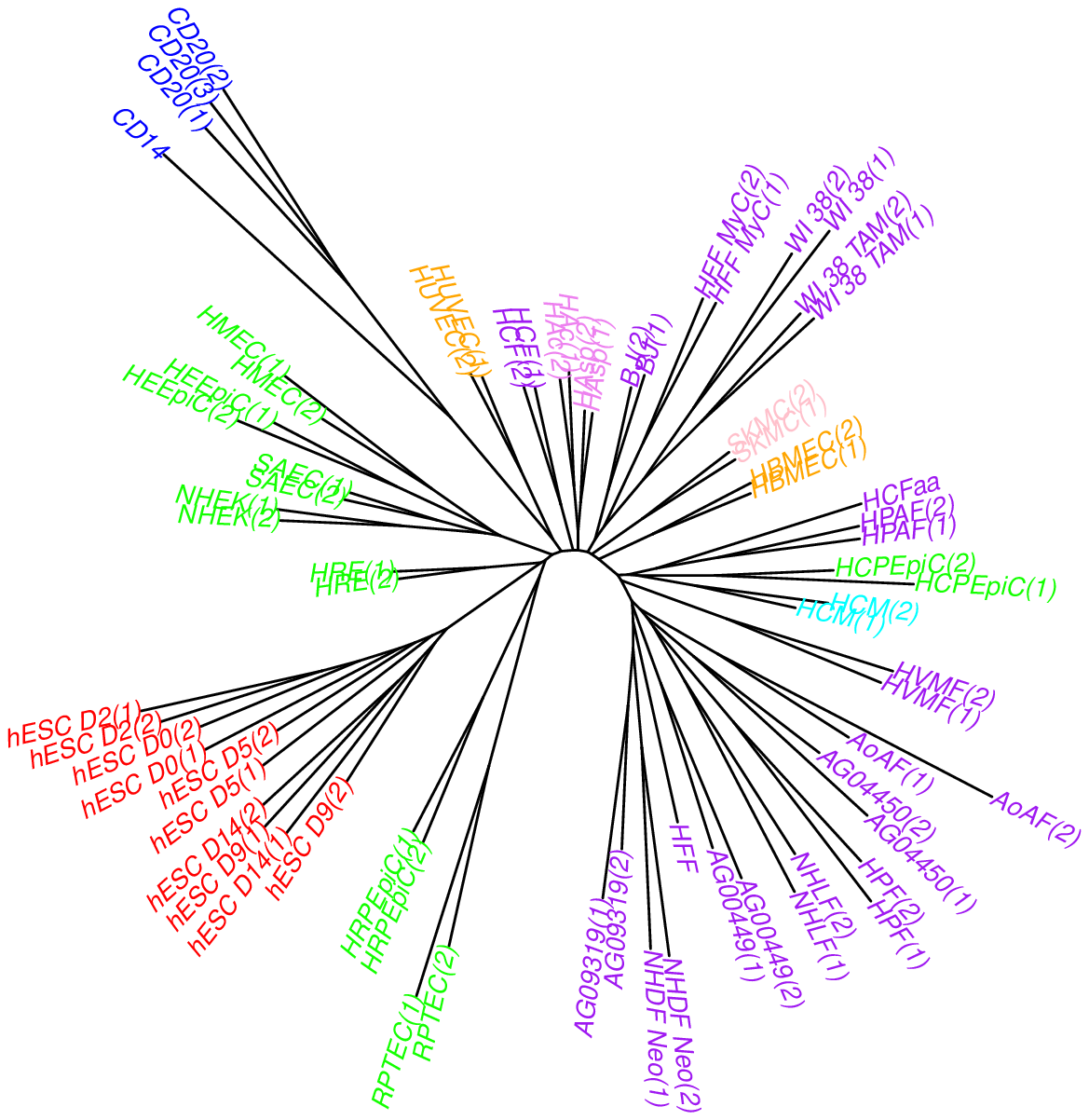}}%
  \hbox to\textwidth{%
	  \hbox to6.9cm{\hfil\scriptsize\it (a) \hfil}\hfil
		  \hbox to6.3cm{\hfil\scriptsize\it (b) \hfil}}%
    \caption{\footnotesize\it Cell-type tree on H3K4me3 data (using all replicates): (a) windowing representation, (b) overlap representation.}%
    \label{fig:H3K4me3_both_rep}%
\end{figure*}

\begin{figure*}[b]%
  \hbox to\textwidth{\hfil
    \includegraphics[width = 6cm]{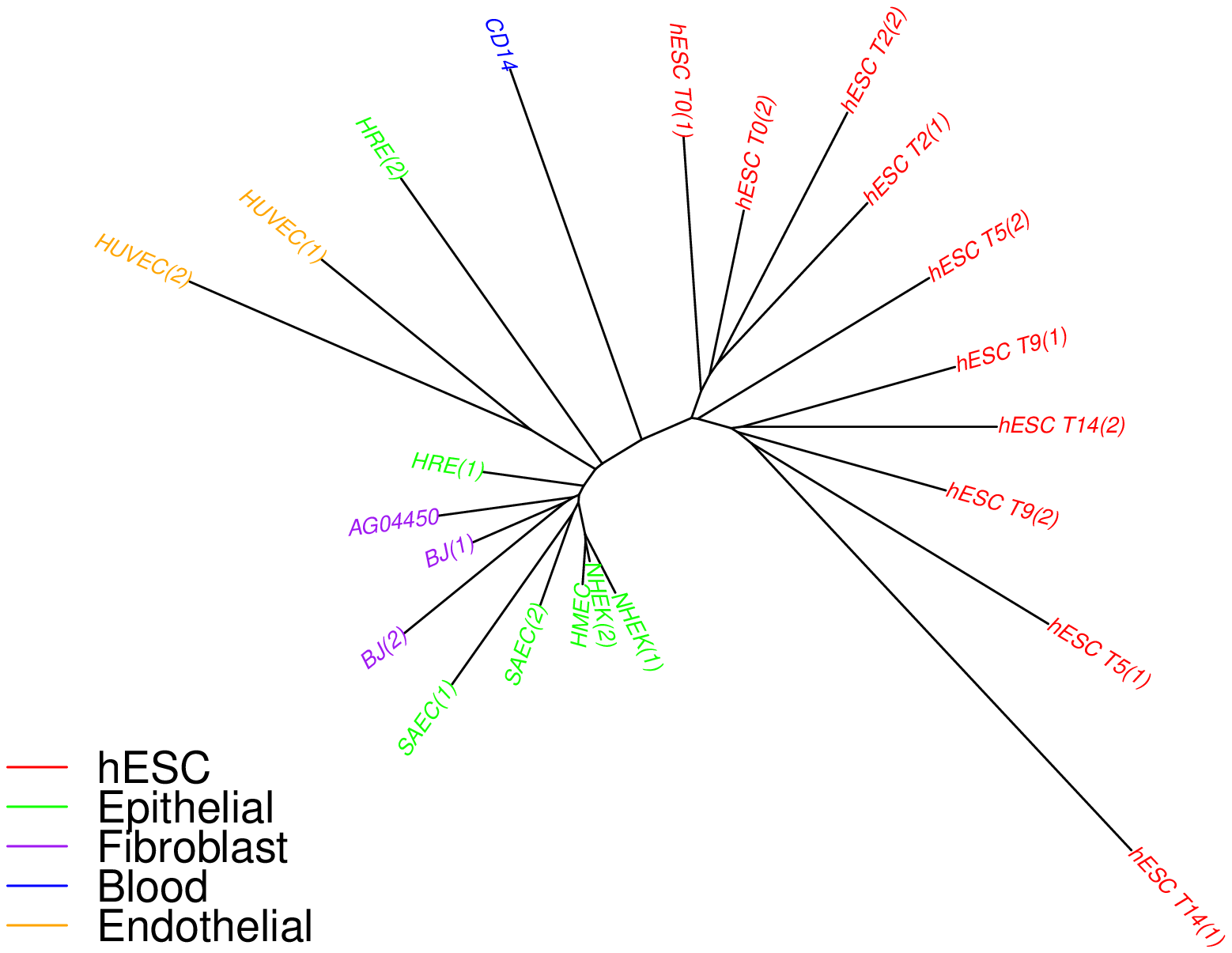}%
  \hfil}%
  \caption{\footnotesize\it Cell-type tree on H3K27me3 data, using all replicates and overlap representation.}%
    \label{H3K27me3}%
\end{figure*}

By bringing replicates into the analysis, we can expect to see a stronger
phylogenetic signal as each replicate adds to the characterization of its
cell type.  In particular, wherever we have two or more replicates, they
should form a tight subtree of their own.  We thus used our replicate
data (two replicates for 33 of the 37 cell types, and three for one type,
for a total of 72 libraries) in the same analysis pipeline.
Fig.~\ref{fig:H3K4me3_both_rep}
shows the differentiation trees obtained using windowing and overlap
representations.
For completeness, we include the same study (in overlap representation
only) on H3K27me3 data in Fig.~\ref{H3K27me3}.
(Finally, the trees obtained using TNT are very similar and not shown.)\ \ 
As expected, almost all replicates are grouped; since we usually have two
replicates, we get a collection of ``cherries" (pairs of leaves)
where we had a single leaf before.  In most cases,
it is now the distance from each leaf in a cherry to their common parent
that is large, indicating that the distance between the two replicates
is quite large---as we can also verify from the distance matrix.
This suggests much noise in the data.  This noise could be
at the level of raw ChIP-Seq data, but also due to the bias of peak-finding
methods used---one expects a general-purpose peak finder to be biased against
false negatives and more tolerant of false positives, but for our application
we would be better served by the inverse bias.  Another reason for
the large distance is the nature of the data: these are biological
replicates, grown in separate cultures, so that many random losses or gains
of histone marks could happen once the cell is differentiated.
Thus it may be that only a few of the mutations in the data
are correlated with cell differentiation. 
Identifying these few mutations would be of high interest, but with just
two replicates we are unlikely to pinpoint them with any accuracy.

Looking again at Table \ref{tab:group}, we see that, using the windowing
representation, the value of $SR$ for the full set of replicates is 0.84
and that here the overlap representation, which is more effective at noise
filtering, yields an $SR$ value of 0.78. This is a significant reduction
and indicates that the long edges are indeed due to noise.
The $PD$ percentage values remain very low for both representations, so
the trees we obtained do represent the data well.
Note that the groupings appear (in the color-coding in the figure)
somewhat better than when we used only one replicate, and the values
in columns 2 through 9 of Table \ref{tab:group} confirm this impression.

\subsection{Using top peaks and masking regions}
\begin{figure*}[b]%
  \hbox to\columnwidth{%
   \includegraphics[width = 6.3cm]{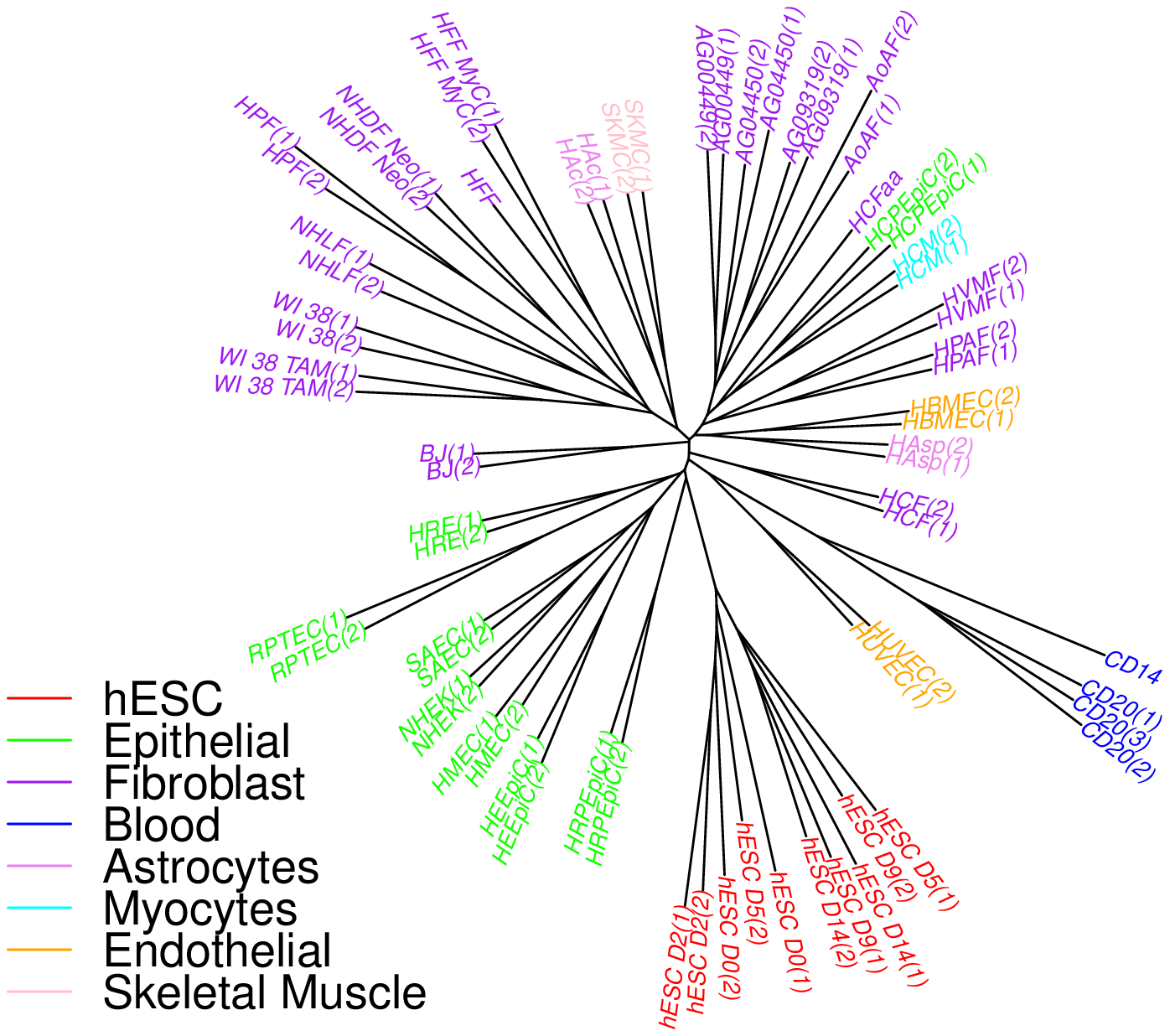}\hfil
   \includegraphics[width = 6.3cm]{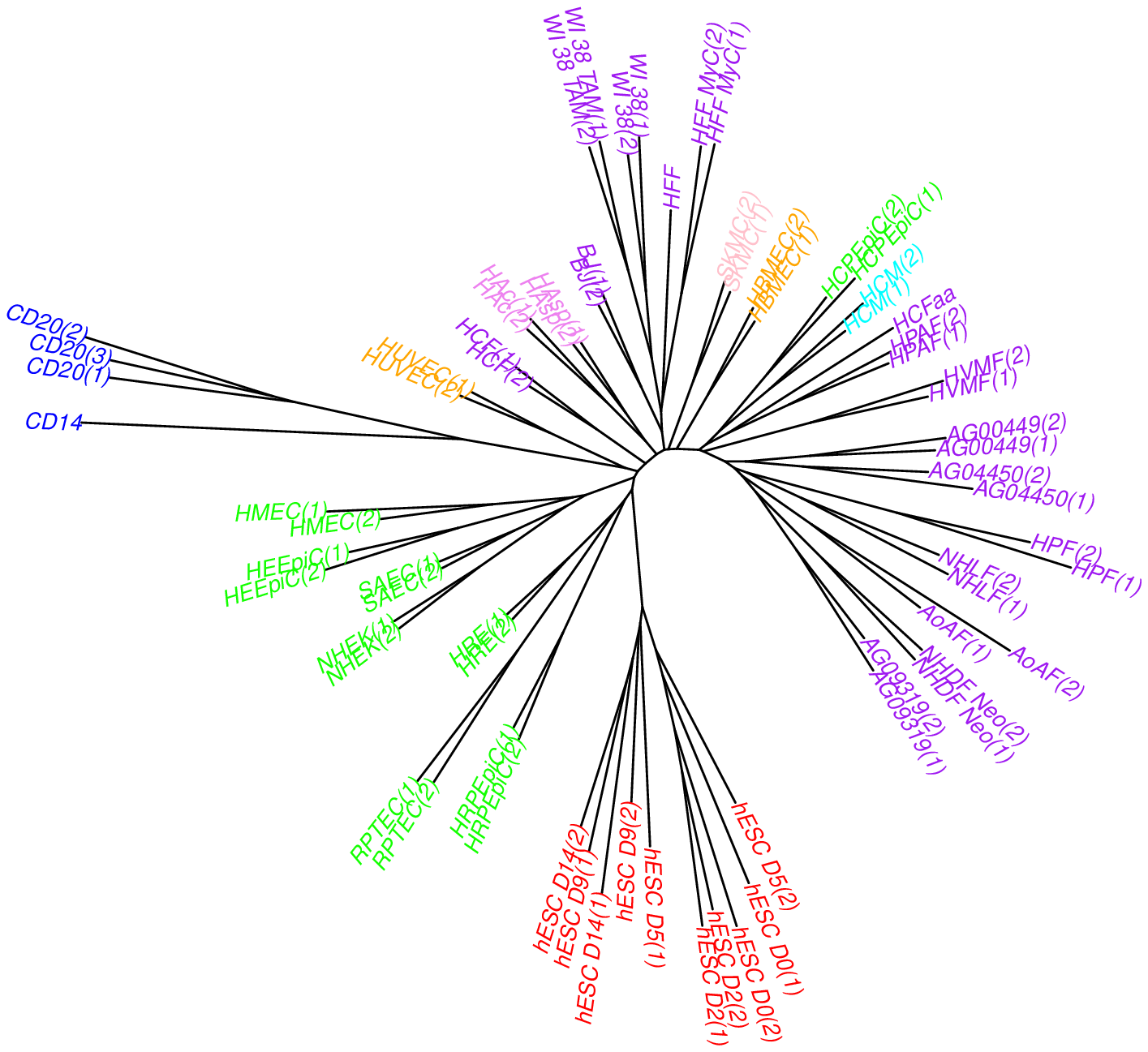}}%
  \hbox to\columnwidth{%
	  \hbox to7cm{\hfil\scriptsize\it (a) \hfil}\hfil
  	  \hbox to7cm{\hfil\scriptsize\it (b) \hfil}}%
    \caption{\footnotesize\it Cell-type tree on H3K4me3 data (using all replicates) on peaks with negative log p-value $\ge$ 10: (a) windowing representation, (b) overlap representation.}%
    \label{fig:H3K4me3_both_rep_top_peaks}%
\end{figure*}

In order to study the nature of the noise, we removed some of the less
robust peaks.  The ENCODE dataset gives a p-value for each peak listed;
we kept only peaks with (negative) log p-values larger than 10.
We kept all replicates and ran the analysis again, with the results
depicted in Fig.~\ref{fig:H3K4me3_both_rep_top_peaks}
The $PD$ percentage values are again very low, so the trees
once again fit the data well.
The improvement looks superficially minor, but we obtained some 
more biologically meaningful clusters with this approach.
For example, in the fibroblast group,
the top two subtrees in Table~\ref{tab:group} changed from (9,4) to (8,5)
when we used only top peaks in the overlap method.  This change occurred
because cell HFF moved from the larger group to the smaller group forming a
subtree with HFF-Myc (which makes more sense as both are foreskin
fibroblast cells).  Such a change could be due to particularly noisy
data for the HFF cells having obscured the relationship before we removed
noisy peaks.  Overall, removing noisy peaks further reduced the $SR$ ratio from
0.78 to 0.74 for the overlap representation and from 0.84 to 0.81 
for the windowing representation.

Another typical noise-reduction procedure, much used in sequence analysis,
is to remove regions that appear to carry little information or to produce
confounding indications---a procedure known as masking.  We devised a 
very simplified version of masking for our problem, for use only with
replicate data, by removing any
region within which at most one library gave a different result (1 instead
of 0 or vice versa) from the others.  In such regions, the presence of
absence of peaks is perfectly conserved across all but one replicate, indicating
the one differing replicate has probably been called wrong.  After
removing such regions, we have somewhat shorter representations, but
follow the same procedure.   The trees returned have exactly the same
topology and so are not shown; the length of edges changed very slightly,
as the $SR$ value decreased from 0.74 down to 0.70 using top peaks
in the overlap representation.

\subsection{A better looking tree}
\begin{figure}[b]%
  \centering
  \includegraphics[height=5.8cm]{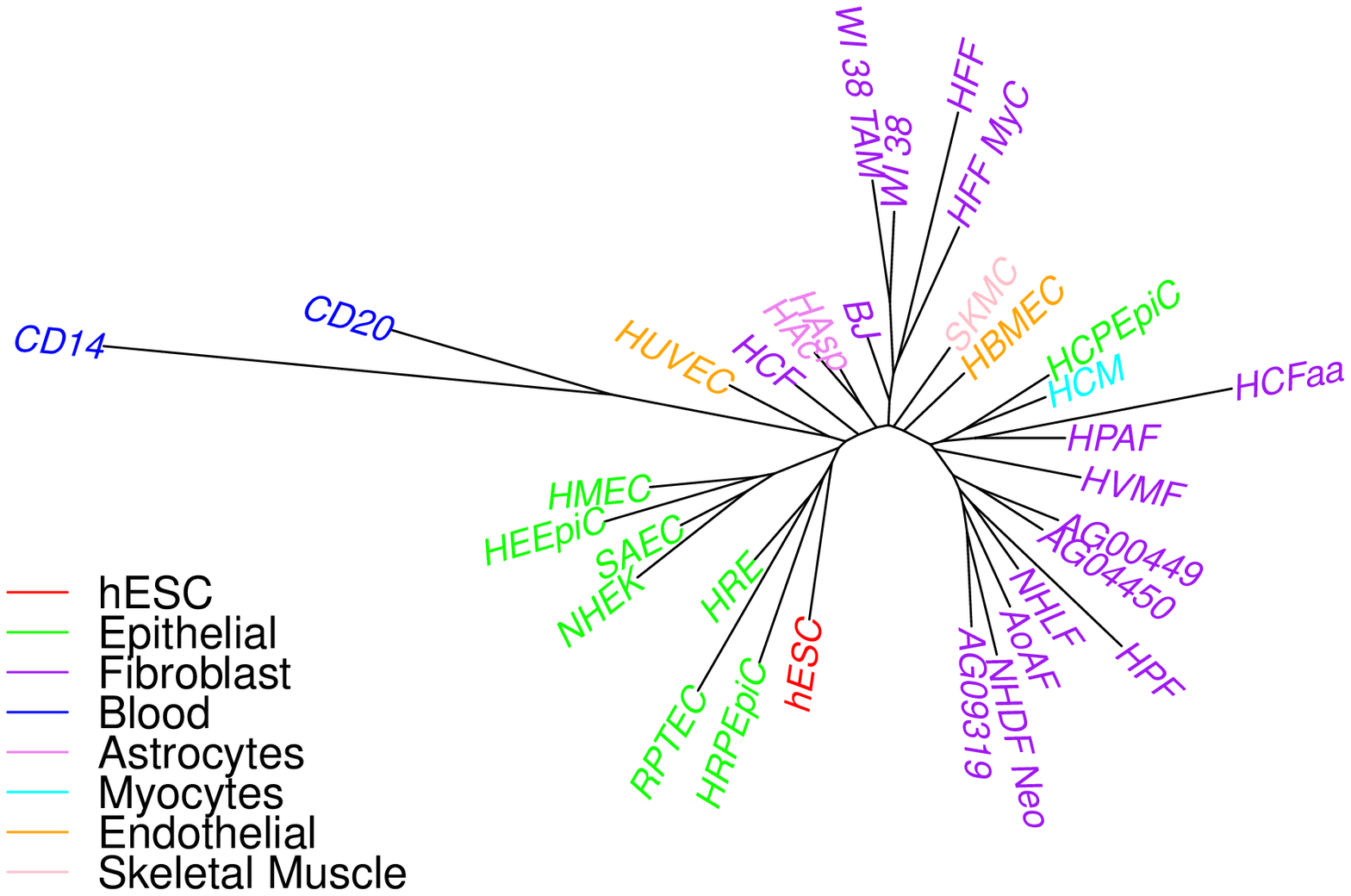}%
  \caption{%
    H3K4me3 data, overlap representation on peaks with negative log p-value $\ge$ 10.
    Replicate leaves are removed and replaced by their parent.}
  \label{fig:H3K4me3_collapse_top_peaks}%
\end{figure}

Barring the addition of many replicates, the $SR$ ratio of 0.70 appears
difficult to reduce and yet remains high.  However, the cherries of replicate
pairs by themselves give an indication of the amount of ``noise" (variation
among individual cells as well as real noise) present in the data.  We can take
that noise out directly by replacing each cherry with its parent, which is
a better representative of the population of this particular cell type than
either of the two leaves.  We carried out this removal on the tree of
Fig.~\ref{fig:H3K4me3_both_rep}(b)
and obtained the tree shown in Fig.~\ref{fig:H3K4me3_collapse_top_peaks}.
Since hESC cells do not form clear pairs, we replaced the entire clade of
hESC cells by their last common ancestor.  The leaves with remaining long
edges are those for which we did not have a replicate (CD14, HCFaa, and HFF).

\section{Conclusions}
We addressed the novel problem of inferring cell-type trees from histone
modification data. We defined methods for representing the peaks as 0/1 vectors
and used these vectors to infer trees.  We obtained very good trees, conforming
closely to expectations and biologically plausible, in spite of the high level
of noise in the data and the very limited number of samples  per cell type.
Our results confirm that histone modification data contain much information
about the history of cell differentiation. 
We carried out a number of experiments to understand the source of the noise,
using replicate data where available, but also devising various noise filters.
Our results show that larger replicate populations are needed to 
infer ancestral nodes, an important step in understanding the process of
differentiation.  Refining models will enable the use of likelihood-based
methods and thus lead to better trees.
Since many histone marks appear independent of cell differentiation,
identifying which marks are connected with the differentiation process is of
significant interest.  Finally, once such marks have been identified,
reconstructing their state in ancestral nodes will enable us to identify 
which regions of the genome play an active role in which steps of
cell differentiation.

\bibliographystyle{splncs03}
\bibliography{diff_tree}

\end{document}